# 200 MeV Ag$^{+15}$ ion irradiation created columnar defects and enhanced critical current density of La-2125 type superconducting thin films

**K. R. Mavani,[a,b] D. S. Rana,[a] S. Rayaprol, R. N. Parmar, D. G. Kuberkar**
*Department of Physics, Saurashtra University, Rajkot – 360005, Gujarat, India*

**Ravi Kumar**
*Materials Science Division, Inter University Accelerator Centre, Aruna Asaf Ali Marg, New Delhi – 110067, India*

**M. Tonouchi**
*Institute of Laser Engineering, Osaka University, 2-6 Yamadaoka, Suita, Osaka 565-0871, Japan*

**J. John and R. Nagarajan**
*Dept. of Condensed Matter Physics and Materials Science, Tata Institute of Fundamental Research, Colaba, Mumbai-400005, India*

## Abstract

We have deposited *c*-axis oriented thin films of La$_{1.5}$Dy$_{0.5}$CaBa$_2$Cu$_5$O$_z$ (La-2125) tetragonal superconductor on LaAlO$_3$ *(001)* by Pulsed Laser Deposition. These films were irradiated with 200 MeV Ag$^{+15}$ ions. Atomic Force Microscopy and Elastic Recoil Detection Analysis indicate that the irradiation has created columnar defects along the entire thickness (2000 Å) of these films. With ion irradiation up to $1\times10^{11}$ ions/cm$^2$, the critical current density (J$_c$(H)) is found to enhance by almost five folds, which is attributed to the augmented flux pinning by the columnar defects. A further increase in irradiation to $1\times10^{12}$ ions/cm$^2$ causes reduction in J$_c$(H) due to distorted morphology of the film. Our work shows that the enhancement in J$_c$ (H) of the irradiated La-2125 film is comparable to that in irradiated RE-123. Also, as the La-2125 type films have greater chemical stability than RE-123 (RE = rare earth ion), La-2125 type superconductors are potential candidates for applications. Interesting to note that there are partial flux jumps observed to occur symmetrically in the magnetic hysteresis of irradiated La-2125 thin films with enhanced J$_c$(H).

a) Present address: Institute of Laser Engineering, Osaka University, Osaka, Japan.

b) Email: mavani-k@ile.osaka-u.ac.jp, krushna1@gmail.com

The perovskite-type $La_{1.5}Ba_{1.5}Cu_3O_z$ is a non-superconducting tetragonal system (space group: *P4/mmm*, no. 123). This system is made superconducting (with superconducting transition temperature, $T_c \sim 78$ K) by suitable hole-doping, such as, $La_{1.5}RE_{0.5}CaBa_2Cu_5O_z$ (RE = rare earth ion) [structure and stoichiometry are described in ref. 1]. It has been shown earlier that such La and Ca substituted tetragonal superconductors have enhanced corrosion resistance compared to RE-123 type orthorhombic superconductors.[2,3] This type of tetragonal superconductors have very stable oxygen stoichiometry and structure with lesser aging effect in the atmosphere, whereas orthorhombic structure and superconducting properties of RE-123 may get altered or even lost due to lesser stability of Oxygen stoichiometry and larger aging effect in the atmosphere. This quality increases the potential for application of the La and Ca based tetragonal superconductors.

The critical current density ($J_c$) is a crucial parameter for the application of any superconductor. When the applied magnetic field is parallel to *c*-axis of the high $T_c$ superconductors (HTSCs), the $J_c$ is particularly low compared to $J_c$ at other orientations and limits the use of the application of the HTSCs.[4] The $J_c$ easily gets influenced by external factors such as impurities, defects, etc.[5] In this regard, as compared to other defects, the columnar defects are known to be more efficient for pinning the flux lines, for enhancing $J_c(H)$, and for shifting up the irreversibility line.[6] The columnar defects have a strong pinning force and when created along the *c*-axis of HTSC, they can increase $J_c$ by boosting up the coupling among vortices in the adjacent $CuO_2$ layers. Thus, one can have improved performance of HTSC with applied magnetic field parallel to *c*- axis. The columnar defects can be introduced by swift ion irradiation,[6] which is a controlled method for inducing defects homogeneously in thin samples. Here, we report that the $J_c$ of La-2125 thin films can be enhanced up to five folds with columnar defects created by heavy ion irradiation.

The thin films of $La_{1.5}RE_{0.5}CaBa_2Cu_5O_z$ type HTSC have note been studied earlier for the heavy ion irradiation induced modifications in $J_c$. With this objective, we synthesized $La_{1.5}Dy_{0.5}CaBa_2Cu_5O_z$ (La-2125) compound, prepared thin films of this compound, irradiated films with Ag ions for introducing columnar defects at an angle perpendicular to $CuO_2$ layers and parallel to $c$-axis and investigated $J_c$ of the films. The 200 MeV $Ag^{+15}$ ions were chosen for irradiation due to the following reasons: i) As per the simulation study by a scientific software SRIM (Stopping and Range of Ion in Matter), 200 MeV $Ag^{+15}$ ions can produce columnar defects through 2000 Å thickness (which is approximately the thickness of our films) of La-2125 films due to high electronic energy loss of the ions, and ii) it was revealed by Elastic Recoil Detection Analysis (ERDA) that the oxygen stoichiometry of La-2125 thin films remains stable during 200 MeV $Ag^{+15}$ ion irradiation.[7]

The synthesis of bulk polycrystalline La-2125 type compounds has been reported in ref. 1. A single-phase $La_{1.5}Dy_{0.5}CaBa_2Cu_5O_z$ was used as target compound during synthesis of thin films using Pulsed Laser Deposition technique. The KrF excimer laser ($\lambda = 248$ nm), with pulse repetition rate of 10 Hz, was used to deposit the La-2125 thin films ($\sim$2000 Å) on single crystal $LaAlO_3$ (*001*) substrates. The energy density was set to 0.9 J/cm$^2$ on the target. The substrate was kept at 4.5 cm distance from the target and at a temperature of 820°C. The O$_2$ partial pressure was maintained at 500 mTorr during the deposition. The X-ray diffraction (XRD) measurements were carried using Philips X-ray diffractomter. The thickness of thin films was measured using a thickness profiler. The resistance was measured as a function of temperature for determining the T$_c$ of the thin films. 200 MeV Ag$^{+15}$ ions were bombarded on several La-2125 thin films, using 15 UD accelerator at Inter University Accelerator Centre, New Delhi, India. The surface morphology of the pristine and irradiated thin films was studied using Atomic Force Microscope (AFM). J$_c$ was determined through magnetic hystereis measurements. In the magnetization vs. magnetic field measurements using SQUID magnetometer (Quantum Design), the magnetic field was applied in a direction perpendicular to Cu-O planes and parallel to irradiation defects, where the pinning force remains maximum.[6]

The analysis of XRD patterns showed that the thin films are single La-2125 phase and highly oriented to *c*-axis on single crystal LaAlO$_3$ (*001*) substrates. As observed from resistivity vs. temperature plot (Fig. 1), these films show the onset of superconducting transition at ~79 K and the T$_c$ ~ 75 K, which is close to the T$_c$ (~78 K) of single-phase bulk La-2125 compound. The surface images of pristine and irradiated thin films are shown in Fig. 2. The irradiated thin films show defects on the surface. The ERDA of La-2125 film showed that the oxygen content of these films is close to the stoichiometry.[7] Furthermore, we found an evidence of columnar defects from ERDA as described in the following way. When 200 MeV Ag$^{+15}$ ions were bombarded on La-2125 thin films, the spectrum of recoiled particles contained Al (Fig. 2d). The only source of Al was LaAlO$_3$ substrates of these thin films. The atoms of Al from the substrates can get recoiled only when Ag ions can pass through the entire thickness of the thin film material and reach up to the substrate. The simulation study (by SRIM), surface morphology (by AFM), and ERDA corroborate well to show that the irradiation has produced columnar defects along the entire thickness (2000 Å) of the films.

The critical current density, $J_c(H)$, was calculated from magnetic hysteresis data using an extended Bean's model.[8] Fig. 3 shows the calculated $J_c(H)$ for pristine and irradiated thin films as function of applied magnetic field parallel to $c$-axis. The calculated $J_c(H)$ for pristine film is around $3.2 \times 10^6$ A/cm$^2$ in presence of very weak magnetic field, which is consistent with the measured $J_c$ ($2.8 \times 10^6$ A/cm$^2$) reported for a similar superconducting thin film in absence of magnetic field.[9] We found (Fig. 3) that $J_c(H)$ of La-2125 thin films enhances with increasing Ag$^{+15}$ ion irradiation up to an ion dose of $1 \times 10^{11}$ ions/cm$^2$. Interestingly, the maximum enhancement in $J_c(H)$ is observed to be almost five folds higher ($1.6 \times 10^7$ A/cm$^2$) compared to $J_c(H)$ of pristine thin films. This large enhancement in $J_c(H)$ of irradiated thin film is attributed to augmented flux pinning by the columnar defects. The enhanced value of $J_c$ is comparable to the $J_c$ of Ag doped RE-123 (for example, ref. 10) and also similar to that of irradiated RE-123 films reported earlier (for examples, refs. 6 and 11).

At an ion dose of $5 \times 10^{11}$ ions/cm$^2$, $J_c(H)$ of the film reduces and shows a crossover with the $J_c(H)$ of pristine thin film (Fig. 3). This is expected as there would be a competition between enhancement of $J_c$ due to flux pinning and decrease in $J_c$ due to damage being caused to the superconducting material. At higher fields, the flux pinning effect dominates and the $J_c$ is more than that of the pristine film. Such a cross over has also been observed in other materials (see for example, ref. 12) where they report a reduction in $T_c$ due to damage to a superconductor. In our case, this is further confirmed by observation of distorted surface morphology of the film due to ion irradiation up to $1 \times 10^{12}$ ions/cm$^2$, where apparently a very large part of the film is affected densely by irradiation (Fig. 2). As a result, the $J_c(H)$ of this film is reduced at all applied magnetic fields up to 50 kOe (Fig. 3).

Another noteworthy feature of La-2125 thin films is that the irradiated films with ion doses $5 \times 10^{10}$ ions/cm$^2$ and $1 \times 10^{11}$ ions/cm$^2$ show kinks in $J_c(H)$ (shown by arrows in Fig. 3). These kinks in $J_c(H)$ correspond to the kinks observed in magnetization of these films. Fig. 4 shows a magnetization hysteresis loop with kinks marked by circles. When the magnetic field sweeping rate was higher during the measurements, a large kink was observed as shown by a larger circle in the figure. This large kink occurs symmetrically in first, third and fifth magnetic field-cycles. Similarly, the occurrence of the small kink is also observed to be symmetric. Similar kinks in magnetization or $J_c(H)$ have been earlier observed for unirradiated superconducting sample[13,14] and also in an irradiated superconducting thin film.[15] A detailed theoretical explanation on such kinks is given by Müller and Andrikidis, taking into account the parameters such as critical magnetic field to destroy superconductivity at specific temperatures, $J_c(H=0)$, $T_c$, specific heat, etc.[13] The type of kinks for these irradiated thin films reflects the partial flux jumps.[16] The possibility of flux jumps enhances when the superconductor is thin, containing well-aligned grains, showing a high $J_c$ or when a high magnetic field sweep-rate is used.[13, 14] All of these conditions apply in the present case, and as a result, we have observed the flux jumps. However, the heat dissipation during the event of kink is not sufficient to de-pin the flux lines and therefore, to show a complete flux jump[15] in these thin films. Importantly, the flux jumps are observed only in those La-2125 films which are irradiated and show enhanced $J_c(H)$. Furthermore, the largest flux jump is observed for the film with highest $J_c(H)$. This may indicate towards a possible correlation between flux jumps, magnetic field, $J_c(H)$ and defects density in such thin films, and requires further investigations.

In summary, we have deposited *c*-axis oriented La-2125 thin films and irradiated them with 200 MeV $Ag^{+15}$ ions. This irradiation produced columnar defects parallel to *c*-axis and perpendicular to $CuO_2$ layers, through the entire thickness of ~2000 Å thin films. An optimal ion dose of $1 \times 10^{11}$ ions/cm$^2$ enhances the critical current density by nearly five folds due to the augmented flux pinning by the columnar defects. This value of $J_c$ is comparable to the $J_c$ obtained in irradiated RE-123. The symmetrically occurring kinks in magnetic hysteresis point towards the partial flux jumps in irradiated La-2125 thin films and a possible correlation of flux jumps with other properties of these films. With the chemical stability of La-2125 being better than that of RE-123, we believe that the La-2125 materials have the potential for applications and deserve to be more intensively investigated.

Authors would like to acknowledge Mr. Fouran Singh for his help during irradiation experiments, Mr. U. D. Vaishnav for his help in AFM measurements and Mr. Nilesh

Kulkarni for the XRD measurements. DGK and KRM thank Inter University Accelerator Centre (formerly NSC), New Delhi, for the financial support through UFUP project.

**Figures:**

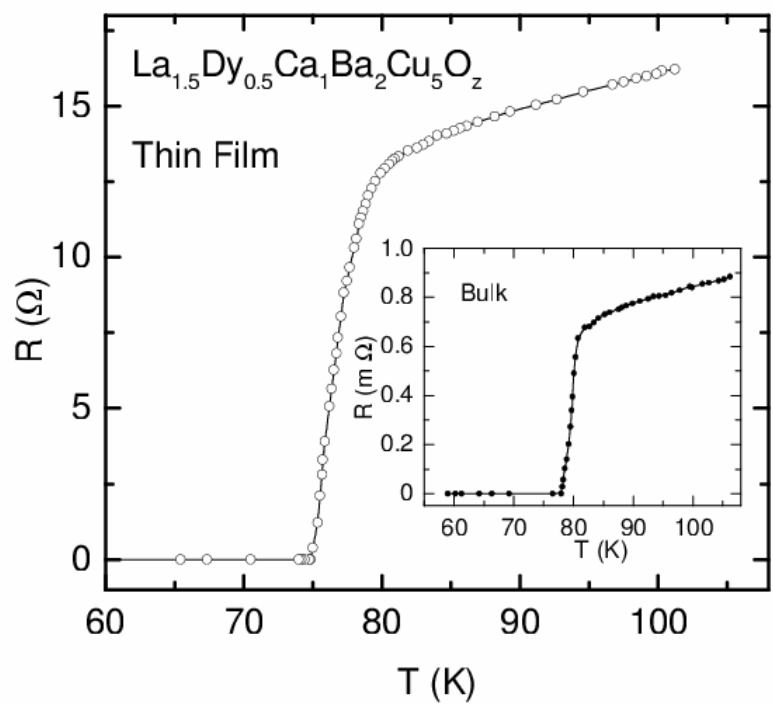

**Fig. 1:** Resistance vs. temperature plot for La-2125 thin film, the inset shows resistance vs. temperature plot for the polycrystalline bulk La-2125.

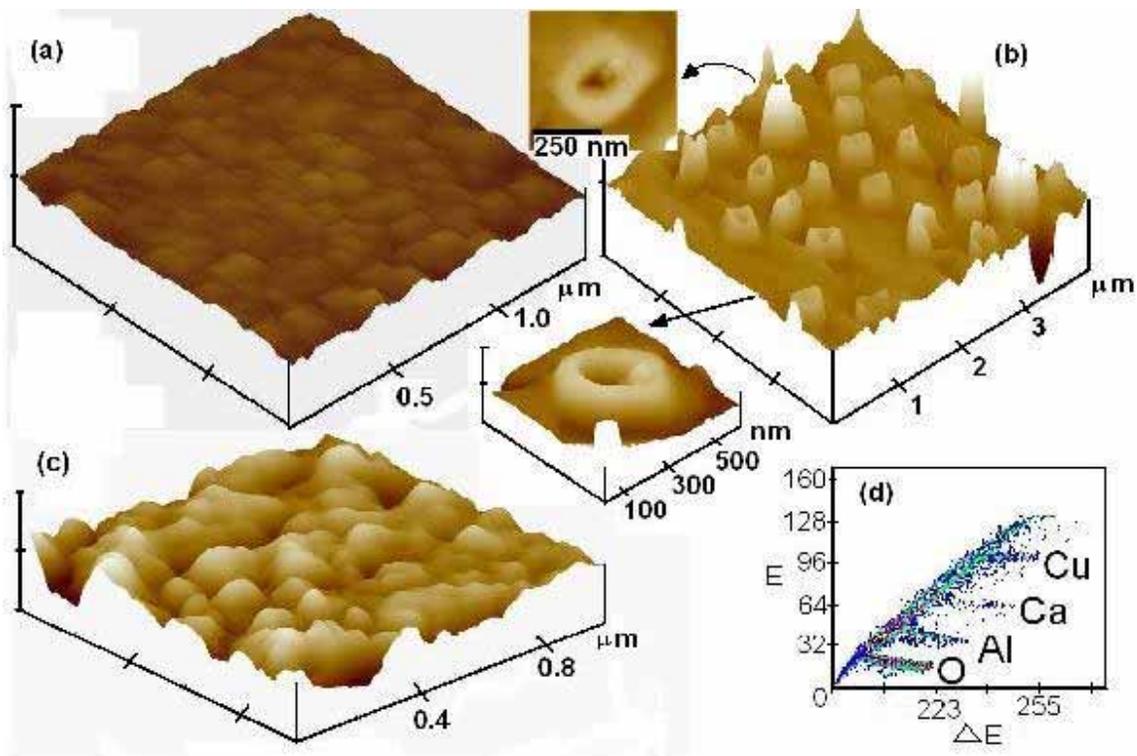

**Fig. 2:** (Color online) Surface morphology, as seen from AFM, of (a) Pristine La-2125 thin film, (b) irradiated La-2125 thin film with optimal ion dose of $1 \times 10^{11}$ ions/cm$^2$; the magnified images (pointed by arrows) show one defect of this film from different angles. (c) Irradiated La-2125 film with highest ion dose of $1 \times 10^{12}$ ions/cm$^2$, and (d) Elastic recoil detection spectrum showing a branch for Al atoms along with Cu, Ca and O atoms, detected by ΔE-E large area position sensitive detector telescope containing total two detectors; where both the axes show channel numbers corresponding to a partial energy loss (ΔE) in a transmission type detector and total energy loss (E).

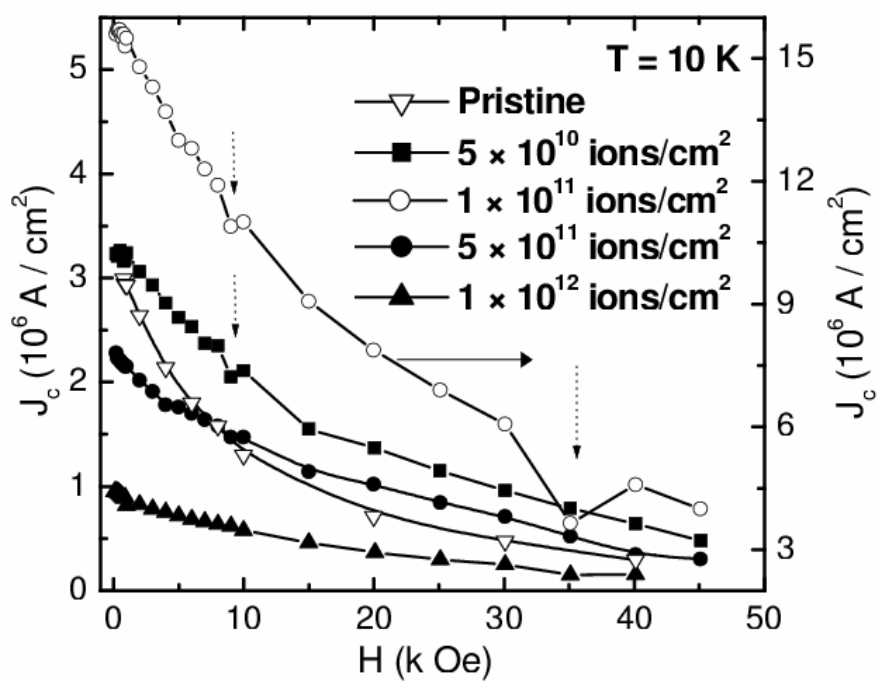

**Fig. 3**: Critical current density ($J_c(H)$) vs. magnetic field plot for pristine and irradiated La-2125 thin films with different ion dose. The vales of $J_c(H)$ (open circles) of the irradiated film with the dose of $1 \times 10^{11}$ ions/cm$^2$ is given on the right hand side of Y-axis. The arrows point towards the kinks in $J_c(H)$.

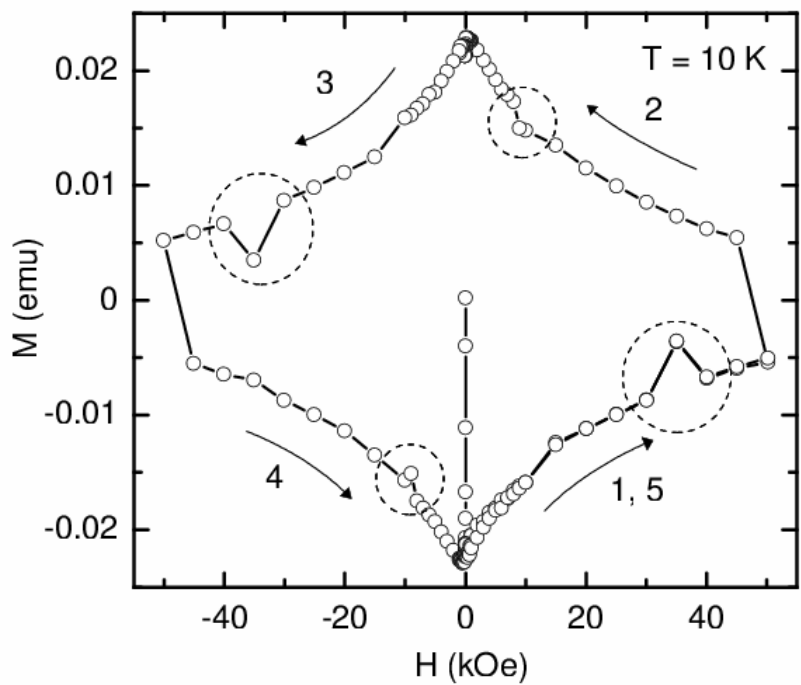

**Fig. 4:** Magnetization hysteresis at 10 K for irradiated La-2125 thin film with an ion dose of $1 \times 10^{11}$ ions/cm$^2$. The symmetrically occurring kinks are marked by circles.

***